\documentstyle{llncs}

\input{psfig}

\begin{document}

\title{Decoherence control for optical qubits}
\author{D. Vitali and P. Tombesi}
\institute{Dipartimento di Matematica e Fisica, Universit\`a di
Camerino \\ via Madonna delle Carceri I-62032 Camerino, Italy}
\maketitle

\begin{abstract}
Photons in cavities have been already used for the realization of 
simple quantum gates [Q.A. Turchette {\it et al.},
Phys. Rev. Lett. {\bf 75}, 4710 (1995)].
We present a method for combatting decoherence 
in this case.
\end{abstract}

\section{Introduction}
Quantum optics is usually concerned with the generation of nonclassical 
states of the electromagnetic field and their experimental detection.
However with the recent rapid progress in the theory of quantum information
processing
the {\it protection} of quantum states and their quantum dynamics also is 
becoming a very important issue.
In fact what makes quantum information processing much more attractive 
than its classical counterpart is its capability of using entangled 
states and of processing generic linear 
superpositions of input states. The entanglement between a pair of 
systems is capable of connecting two observers separated by a space-like 
interval, it can neither be copied nor eavesdropped on without 
disturbance, nor can it be used by itself to send a classical message 
\cite{bennet}.
The possibility of using linear superposition states has given rise to 
quantum computation, which is essentially equivalent to have massive 
parallel computation \cite{eke}. 
However all these applications crucially 
rely on the possibility of maintaining quantum coherence, that is, a 
defined phase relationship between the different components of linear 
superposition states, over long distances and for long times. 
This means that one has to minimize as much as possible the effects of 
the interaction of the quantum system with its environment and, in 
particular, decoherence, i.e., the rapid destruction of the
phase relation between two quantum states of a system caused by the
entanglement of these two states with two different states of the
environment \cite{zur,leg1}.

Quantum 
optics is a natural candidate for the experimental implementation of 
quantum information processing systems, 
thanks to the recent achievements in the manipulation of 
single atoms, ions and single cavity modes. In fact two quantum gates 
have been already demonstrated \cite{turchette,wine1} in quantum 
optical systems and 
it would be very important to develop strategies capable of {\it 
controlling the decoherence} in experimental situations such as 
those described in Refs.~\cite{turchette,wine1}.

In this paper we propose a simple physical way to control decoherence
and protect a given quantum state against the destructive effects of the 
interaction with the environment: applying an appropriate feedback.
The feedback scheme considered here 
has a quantum nature, since it is based on the injection of an 
appropriately prepared atom in the cavity and some preliminary aspects of the
scheme, and its 
performance, have been  described in Refs.~\cite{prlno,jmo}

\section{A feedback loop for optical cavities}

Applying a feedback loop to a quantum system means subjecting it to a 
series of measurements and then using the result of these 
measurements to modify the dynamics of the system. 
Very often the system is continuously monitored and 
the associated feedback scheme provides a continuous control of the 
quantum dynamics. An example is the measurement of an optical 
field mode, such as photodetection and homodyne measurements, and for 
these cases, Wiseman and Milburn have developed a quantum theory of 
continuous feedback \cite{feed}. This theory has been applied in 
Refs.~\cite{noi} to show that homodyne-mediated feedback
can be used to slow down the decoherence of a Schr\"odinger 
cat state in an optical cavity. 

Here we propose a different feedback scheme, 
based on direct photodection rather than homodyne 
detection. The idea is very simple: whenever the cavity looses a 
photon, a feedback loop supplies the cavity mode with another 
photon, through the injection of an appropriately prepared atom.
This kind of feedback is naturally suggested by the quantum 
trajectory picture of a decaying cavity field \cite{Carmichael}, in which 
time evolution is driven by the non-unitary evolution operator 
$\exp\{-\gamma t a^{\dagger}a/2\}$ interrupted at random times by an 
instantaneous jump describing the loss of a photon. 
The proposed feedback almost instantaneously ``cures'' the effect 
of a quantum jump and is able therefore to minimize the destructive effects of 
dissipation on the quantum state of the cavity mode.

In more general terms, the 
application of a feedback loop modifies the master 
equation of the system and therefore it is equivalent to an effective 
modification of the dissipative environment of the cavity field. For 
example, Ref.~\cite{squee} shows that a squeezed bath \cite{qnoise} 
can be simulated by the application of a feedback loop based on a 
quantum non-demolition (QND) measurement of a quadrature of a cavity 
mode. In other words, feedback is the main tool for realizing, 
in the optical domain, the so called ``quantum reservoir 
engineering'' \cite{poyatos}.  

The master equation for continuous feedback has been
derived by Wiseman and Milburn 
\cite{feed}, and, in the case of perfect detection via a single 
loss source, is given by
\begin{equation}
\dot{\rho }= \gamma \Phi (a \rho a^{\dagger}) -\frac{\gamma 
}{2}a^{\dagger}a \rho -\frac{\gamma }{2} \rho a^{\dagger}a \;,
\label{feedeq}
\end{equation}
where $\gamma $ is the cavity decay rate and $\Phi (\rho )$ is a generic 
superoperator describing the 
effect of the feedback atom  on the cavity state $\rho$.
Eq.~(\ref{feedeq}) assumes perfect detection, i.e., 
all the photons leaving the cavity are absorbed by a unit-efficiency
photodetector and trigger the cavity loop. It is practically impossible to
realize such an ideal situation and therefore it is more realistic to 
generalize this feedback master equation to the situation where
only a fraction  $\eta < 1$ of the photons leaking out of the cavity
is actually detected and switches on the atomic injector. It is 
immediate to see that (\ref{feedeq}) generalizes to
\begin{equation}
\dot{\rho }= \eta \gamma \Phi (a \rho a^{\dagger}) 
+(1-\eta) \gamma a \rho a^{\dagger}-\frac{\gamma 
}{2}a^{\dagger}a \rho -\frac{\gamma }{2} \rho a^{\dagger}a \;. 
\label{feedeq2}
\end{equation}

Now, we have to determine the action of the feedback atom on the cavity 
field $\Phi (\rho )$; this atom has to release exactly one photon 
in the cavity, possibly regardless of the field state in the cavity. In the 
optical domain this could be realized using
{\it adiabatic transfer of Zeeman coherence} \cite{adia}.

\subsection{Adiabatic passage in a three level $\Lambda $ atom}

A scheme based on the adiabatic passage of an atom with Zeeman 
substructure through overlapping cavity and laser fields has been 
proposed \cite{adia} for the generation of linear superpositions of Fock 
states in optical cavities. This technique allows for coherent 
superpositions of atomic ground state Zeeman sublevels to be ``mapped'' 
directly onto coherent superpositions of cavity-mode number states.
If one applies this scheme in the simplest case of a three-level $\Lambda 
$ atom one obtains just the feedback superoperator we are looking for, that is
\begin{equation}
\Phi (\rho ) = 
a^{\dagger} (aa^{\dagger})^{-1/2} \rho (aa^{\dagger})^{-1/2} a \;,
\label{adiafi} 
\end{equation}
corresponding to the feedback atom releasing exactly one photon into the cavity,
regardless the state of the field. 

\begin{figure}
\centerline{\psfig{figure=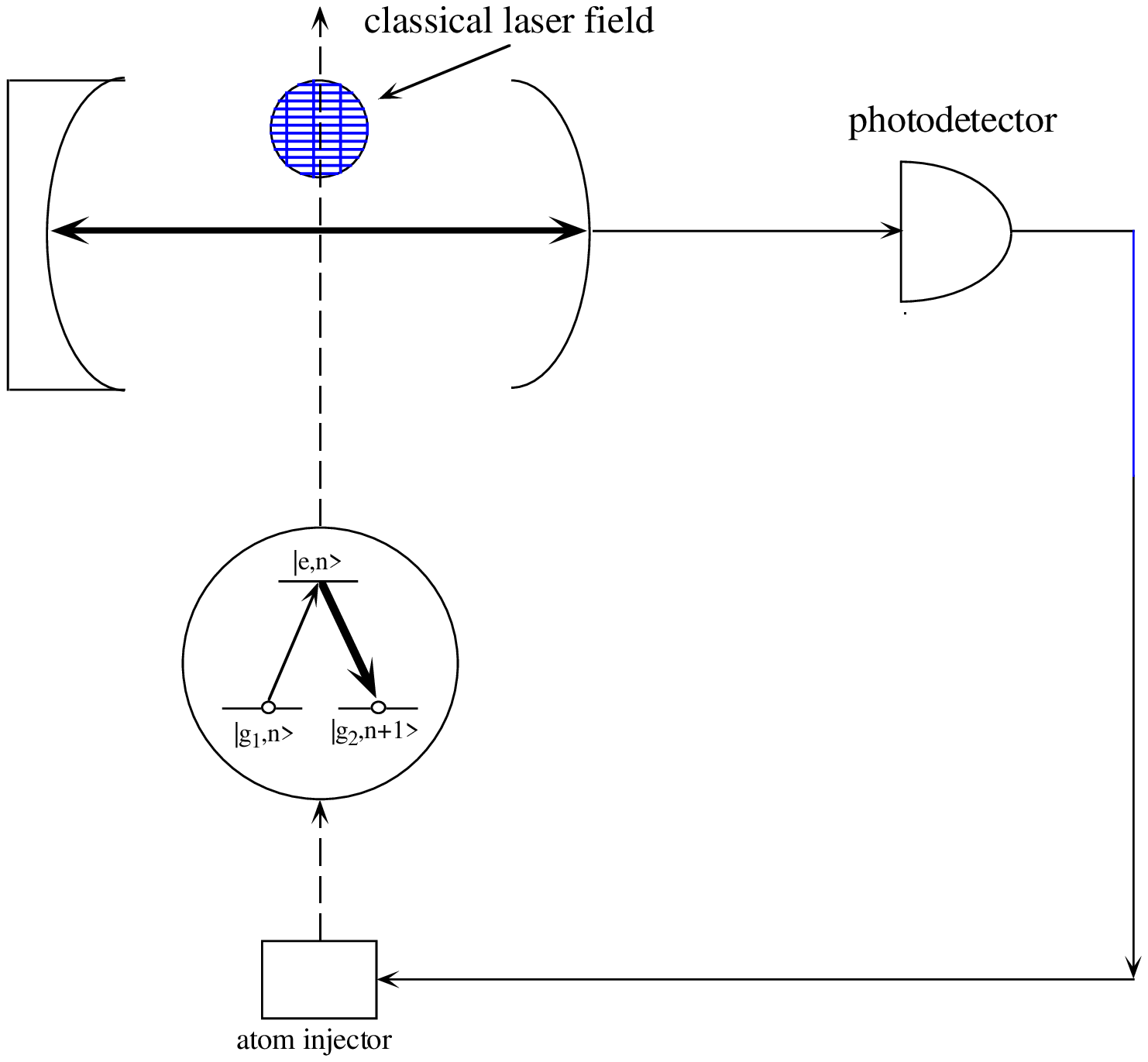,width=12cm}}
\caption{Schematic diagram of the photodetection-mediated feedback scheme
proposed for optical cavities, together with the appropriate atomic 
configuration for the adiabatic transfer.}
\label{appa1}
\end{figure}

To see this, let us consider
a three level $\Lambda$ atom with two ground states $|g_{1}\rangle$
and $|g_{2}\rangle$, coupled to the excited state $|e\rangle $ 
via, respectively, a classical laser field $\Omega (t)$ of frequency 
$\omega _{L}$, and a cavity field mode of frequency $\omega $. The 
corresponding Hamiltonian is
\begin{eqnarray}
&& H(t)= \hbar \omega a^{\dagger}a +\hbar \omega_{eg} |e\rangle \langle e|
-i\hbar g(t) \left(|e\rangle \langle g_{2}| a-|g_{2}\rangle \langle e| 
a^{\dagger}\right)  \nonumber  \\
&&+ i \hbar \Omega(t) \left(|e\rangle \langle 
g_{1}|e^{-i\omega _{L} t}-|g_{1}\rangle \langle e| e^{i \omega _{L} t}\right)
\;.
\label{jcm}
\end{eqnarray}
The time dependence of $\Omega(t)$ and $g(t)$ is provided by the 
motion of the atom across the laser and cavity profiles. This Hamiltonian 
couples only states within the three-dimensional manifold spanned by 
$|g_{1},n\rangle ,\,|e,n\rangle,\,|g_{2},n+1\rangle $, where $n$ denotes 
a Fock state of the cavity mode. Of particular interest within this 
manifold is the eigenstate corresponding to the adiabatic energy 
eigenvalue (in the frame rotating at the frequency $\omega $) 
$E_{n}=n\hbar \omega$,
\begin{equation}  
|E_{n}(t)\rangle = \frac{g(t) \sqrt{n+1}|g_{1},n\rangle  +\Omega(t) 
|g_{2},n+1 \rangle }{\sqrt{\Omega ^{2}(t)+(n+1) g^{2}(t)}}
\end{equation}
which does not contain any contribution from the excited state and for 
this reason is called the ``dark state''. This eigenstate exhibits the 
following asymptotic behavior as a function of time
\begin{equation}
|E_{n}\rangle \rightarrow \left\{\matrix{
                |g_{1},n\rangle & {\rm for} & \Omega(t)/g(t) \rightarrow 0  \cr
                |g_{2},n+1\rangle & {\rm for} & g(t)/\Omega(t) \rightarrow 0  \cr} \right.
\label{asymp}
\end{equation}
Now, according to the adiabatic theorem \cite{messiah}, 
when the evolution from time $t_{0}$ to time $t_{1}$ is sufficiently slow, a 
system starting from an eigenstate of $H(t_{0})$ will pass into the 
corresponding eigenstate of $H(t_{1})$ that derives from it by continuity.
This means that if the atom crossing is such that adiabaticity is satisfied,
when the atom enters the interaction region in the ground state 
$|g_{1}\rangle$, the following adiabatic transformation 
of the atom-cavity system state takes place
\begin{eqnarray}
\label{reali}
&& |g_{1}\rangle \langle g_{1}| \otimes \sum_{n,m}\rho _{n,m} 
|n\rangle \langle m| \\
&& \rightarrow |g_{2}\rangle \langle g_{2}| \otimes 
\sum_{n,m}\rho _{n,m} |n+1\rangle \langle m+1| \nonumber \\
&& = |g_{2}\rangle \langle g_{2}| \otimes 
a^{\dagger} (aa^{\dagger})^{-1/2} \rho (aa^{\dagger})^{-1/2} a \nonumber \;.
\end{eqnarray}
Roughly speaking, this transformation amounts to a single photon 
transfer from the classical laser field to the quantized cavity mode 
realized by the crossing atom, provided that a counterintuitive 
pulse sequence in which the classical laser field $\Omega (t)$ is 
time-delayed with respect to $g(t)$ is applied. Figure \ref{appa1} 
shows a simple diagram of the feedback scheme, together with the 
appropriate atomic configuration, cavity and laser field profiles 
needed for the adiabatic transformation considered. 

The quantitative conditions under which adiabaticity is satisfied are 
obtained from the requirement that the transition from the dark state 
$|E_{n}(t)\rangle $ to the other states be very small. One obtains 
\cite{adia,chim}
\begin{equation}
\Omega_{max},g_{max} \gg T_{cross}^{-1}  \;,
\end{equation}
where $T_{cross}$ is the cavity crossing time and $\Omega_{max},g_{max}$ 
are the two peak intensities. 

The above arguments completely neglect dissipative effects due to cavity 
losses and atomic spontaneous emission. For example, cavity dissipation 
couples a given manifold $|g_{1},n\rangle ,\,|e,n\rangle,\,|g_{2},n+1\rangle $
with those with a smaller number of photons. Since ideal adiabatic 
transfer occurs when the passage involves a single manifold, optimization 
is obtained when the photon leakage through the cavity is 
negligible during the atomic crossing, that is
\begin{equation}
T_{cross}^{-1} \gg  \bar{n} \gamma \;,
\end{equation}
where $\bar{n}$ is mean number of photons in the cavity. On the contrary, 
the technique of adiabatic passage is robust against the effects of 
spontaneous emission as, in principle, the excited atomic state 
$|e\rangle$ is never populated. Of course, in practice some fraction of 
the population does reach the excited state and hence large values of 
$g_{max}$ and $\Omega _{max}$ relative to the spontaneous emission rate 
$\gamma _{e}$ are desirable.
To summarize, the quantitative conditions for a practical realization of 
the adiabatic transformation (\ref{reali}) are
\begin{equation}
\Omega_{max},g_{max} \gg T_{cross}^{-1} \gg \bar{n} \gamma, 
\gamma_{e} \;,
\label{condi}
\end{equation}
which, as pointed out in \cite{adia}, could be realized in optical cavity QED
experiments. 

We note that when the adiabaticity conditions (\ref{condi}) are 
satisfied, then also the Markovian assumptions at the basis of the feedback 
master equation (\ref{feedeq2}) are automatically justified.
In fact, the continuous feedback theory of Ref.~\cite{feed} 
is a Markovian theory derived assuming that
the delay time associated to the 
feedback loop can be neglected with respect to the typical 
timescale of the cavity mode dynamics. In the present 
scheme the feedback delay time is due to the electronic trasmission time of the 
detection signal and, most importantly, by the interaction time $T_{cross} $ 
of the atoms with the field, while the typical timescale of the cavity field 
dynamics is $1/\gamma \bar{n}$. Therefore, the inequality on the right of
Eq.~(\ref{condi}) is essentially the condition for the validity of the Markovian
approximation and this {\it a posteriori} justifies our use of the Markovian 
feedback master equation (\ref{feedeq2}) from the beginning. 

\subsection{Properties of the adiabatic transfer feedback model}

When we insert the explicit expression (\ref{adiafi}) of the feedback 
superoperator into Eq.~(\ref{feedeq2}), the feedback master equation 
can be rewritten in the more transparent form 
\begin{equation}
\dot{\rho }= \frac{(1-\eta) \gamma}{2}\left(2 a \rho a^{\dagger}-
a^{\dagger}a \rho - \rho a^{\dagger}a \right)
-\frac{\eta \gamma}{2} \left [ \sqrt{\hat{n}},\left[\sqrt{\hat{n}},
\rho \right] \right]
\label{sqroot} 
\end{equation}
that is, a standard vacuum bath master equation with 
effective damping coefficient $(1-\eta) \gamma $ plus an unconventional 
phase diffusion term, in which the photon number operator is replaced by 
its square root and which can be called ``square root of phase diffusion''.

In the ideal case $\eta=1$, vacuum damping vanishes
and only the unconventional phase diffusion survives. As shown in 
Ref.~\cite{wise}, this is equivalent to say that ideal photodetection 
feedback is able to transform standard photodetection into a quantum 
non-demolition (QND) measurement of the photon number. In this ideal case,
a generic Fock state $|n\rangle$ is obviously preserved for an 
infinite time, since each photon lost by the cavity triggers the feedback 
loop which, in a negligible time, is able to give the photon back through 
adiabatic transfer. However, the photon injected by feedback 
has no phase relationship with the photons already present in the cavity 
and, as shown by (\ref{sqroot}), this results in phase diffusion. 
This means that feedback does not guarantee perfect state protection for 
a generic {\it superposition of number states}, even in 
the ideal condition $\eta =1$. In fact in this case, only the diagonal matrix 
elements in the Fock basis of the initial pure state are perfectly conserved, 
while the off-diagonal ones always decay to zero, ultimately leading 
to a phase-invariant state. However this does not mean that the 
proposed feedback scheme is good for preserving number states only, 
because the unconventional ``square-root of phase 
diffusion'' is much slower than the conventional one (described by a 
double commutator with the number operator). 

In fact the time evolution of a generic density matrix element in the 
case of feedback with ideal photodetection $\eta =1$ is
\begin{equation}
\rho _{n,m}(t)=\exp\left\{-\frac{\gamma 
t}{2}\left(\sqrt{n}-\sqrt{m}\right)^{2}\right\} \rho _{n,m}(0) \;,
\label{sqrt2}
\end{equation}
while the corresponding evolution in the presence of 
standard phase diffusion is
\begin{equation}
\rho _{n,m}(t)=\exp\left\{-\frac{\gamma 
t}{2}\left(n-m\right)^{2}\right\} \rho _{n,m}(0) \;.
\label{phadif}
\end{equation}

Since 
\begin{equation}
(n-m)^{2} \geq \left(\sqrt{n}-\sqrt{m}\right)^{2} = 
\frac{(n-m)^{2}}{\left(\sqrt{n}+\sqrt{m}\right)^{2}} \;\;\;\;\; 
\forall\;n,m \,
\label{ineq}
\end{equation}
each off-diagonal matrix element decays slower in the square root 
case and this means that the feedback-induced unconventional phase 
diffusion is slower than the conventional one. 

A semiclassical estimation 
of the diffusion constant can be obtained from 
the time evolution of the mean coherent amplitude $\langle 
a(t) \rangle $. In fact, phase diffusion causes a decay of 
this amplitude as the phase spreads around  $2\pi $, even if the 
photon number is conserved.  In the presence of ordinary phase diffusion the
amplitude decays at the rate $\gamma /2$; in fact 
\begin{equation}
\langle a(t) \rangle = {\rm Tr}\left\{a \rho(t)\right\}=
\sum _{n=0}^{\infty}\sqrt{n+1} \rho_{n+1,n}(t) \;,
\label{amedio}
\end{equation}
and using Eq.~(\ref{phadif}) one gets
$$
\langle a(t) \rangle =e^{-\gamma t/2} \langle a(0) \rangle \;.
$$
In the case of the square root of phase diffusion,
Eqs.~(\ref{sqrt2}) and (\ref{amedio}) instead yield
\begin{equation}
\langle a(t) \rangle = {\rm Tr}\left\{a(t) \rho(0)\right\} \;,
\label{amedisq}
\end{equation}
where the Heisenberg-like time evolved amplitude operator $a(t)$ is 
given by
\begin{equation}
a(t) = \exp\left\{-\frac{\gamma t}{2}\left(\sqrt{a a 
^{\dagger}}-\sqrt{a^{\dagger} a}\right)^{2}\right\} a \;.
\label{amediti}
\end{equation}

In the semiclassical limit it is reasonable to assume
a complete factorization of averages, so to get
\begin{equation}
\langle a(t) \rangle = \exp\left\{-\frac{\gamma 
t}{2}\left(\sqrt{\bar{n}+1}-\sqrt{\bar{n}}\right)^{2}\right\} 
\langle a(0) \rangle \;,
\label{amediti2}
\end{equation}
which, in the limit of large mean photon number $\bar{n}$, yields 
\begin{equation}
\langle a(t) \rangle = \exp\left\{-\frac{\gamma 
t}{8 \bar{n}}\right\} \langle a(0) \rangle \;.
\label{amediti3}
\end{equation}
This slowing down of phase diffusion (similar to that taking place 
in a laser well above threshold) means that, 
when the feedback efficiency $\eta $ is not too 
low, the ``lifetime'' of generic pure quantum states of
the cavity field can be significantly increased with respect to the 
standard case with no feedback (see Eq.~\ref{sqroot}). 

\section{Optical feedback scheme for the protection of qubits}

Photon states are known to retain their phase coherence over considerable 
distances and for long times and for this reason high-Q optical cavities 
have been proposed as a promising example for the realization of simple 
quantum circuits for quantum information processing. To act as an 
information carrying quantum state, the electromagnetic fields must 
consist of a superposition of few distinguishable states. The most 
straightforward choice is to consider the superposition of the vacuum and 
the one photon state $\alpha |0\rangle +\beta |1\rangle $. However it is easy 
to understand that this is not convenient because any interaction 
coupling $|0\rangle $ and $|1\rangle $ also couples $|1\rangle $ with 
states with more photons and this leads to information losses. Moreover 
the vacuum state is not easy to observe because it cannot be 
distinguished from a failed detection of the one photon state. A more 
convenient and natural choice is {\it polarization coding}, i.e., using 
two degenerate polarized modes and qubits of the following form
\begin{equation}
|\psi \rangle = \left(\alpha a_{+}^{\dagger}+\beta a_{-}^{\dagger}\right) 
|0\rangle = \alpha |0,1\rangle + \beta |1,0\rangle \;,
\label{twoqub}
\end{equation}
in which one photon is shared by the two modes \cite{sten}.
In fact this is a ``natural'' two-state system, in which the two basis 
states can be easily distinguished with polarization measurements; 
moreover they can be easily transformed into each other using polarizers.

Polarization coding has been already employed in one of the few 
experimental realization of a quantum gate, the quantum phase gate 
realized at Caltech \cite{turchette}. This experiment has demonstrated 
conditional quantum dynamics between two frequency-distinct fields
in a high-finesse optical cavity. The implementation of this gate 
employs two single-photon pulses with frequency separation large 
compared to the individual bandwidth, and whose internal state is 
specified by the circular polarization basis as in (\ref{twoqub}).
The conditional dynamics between the two fields is obtained 
through an effective strong 
Kerr-type nonlinearity provided by a beam of cesium atoms. 
This conditional dynamics of the quantum gate has to be unitary with a 
high degree of 
accuracy during the operation time, i.e., decoherence has to be negligible; 
the experiment of Ref.~\cite{turchette} has 
been performed in the bad cavity limit, in which the main dissipative 
effects and main source of decoherence is
the photon leakage outside the cavity. It is therefore quite natural 
to see if the atomic feedback scheme described in detail above 
is able to protect the ``flying'' qubits of Ref.~\cite{turchette}.
To be more specific, we shall not be concerned with the protection of the 
quantum gate dynamics, but we shall focus on a simpler but still 
important problem: protecting an unknown input state for the longest 
possible time against decoherence.
We shall therefore consider a single qubit, i.e., a single frequency
whose internal state is specified by the polarization.

One has to apply an adiabatic transfer feedback loop 
as that of Fig.~1 to each polarized mode independently. This can be done 
using polarization-sensitive detectors (for example a polarized beam 
splitter and two detectors) and two classical laser fields 
with opposite circular polarization. In this way one has two similar
feedback loops where one polarized mode is involved
in the transition $|g_{1}\rangle \rightarrow |g_{2}\rangle $, 
and
the other mode participates to the reversed transition. In this way each mode 
gets a photon with the right polarization. A schematic description 
of the scheme is given by Fig.~2.
\begin{figure}
\centerline{\psfig{figure=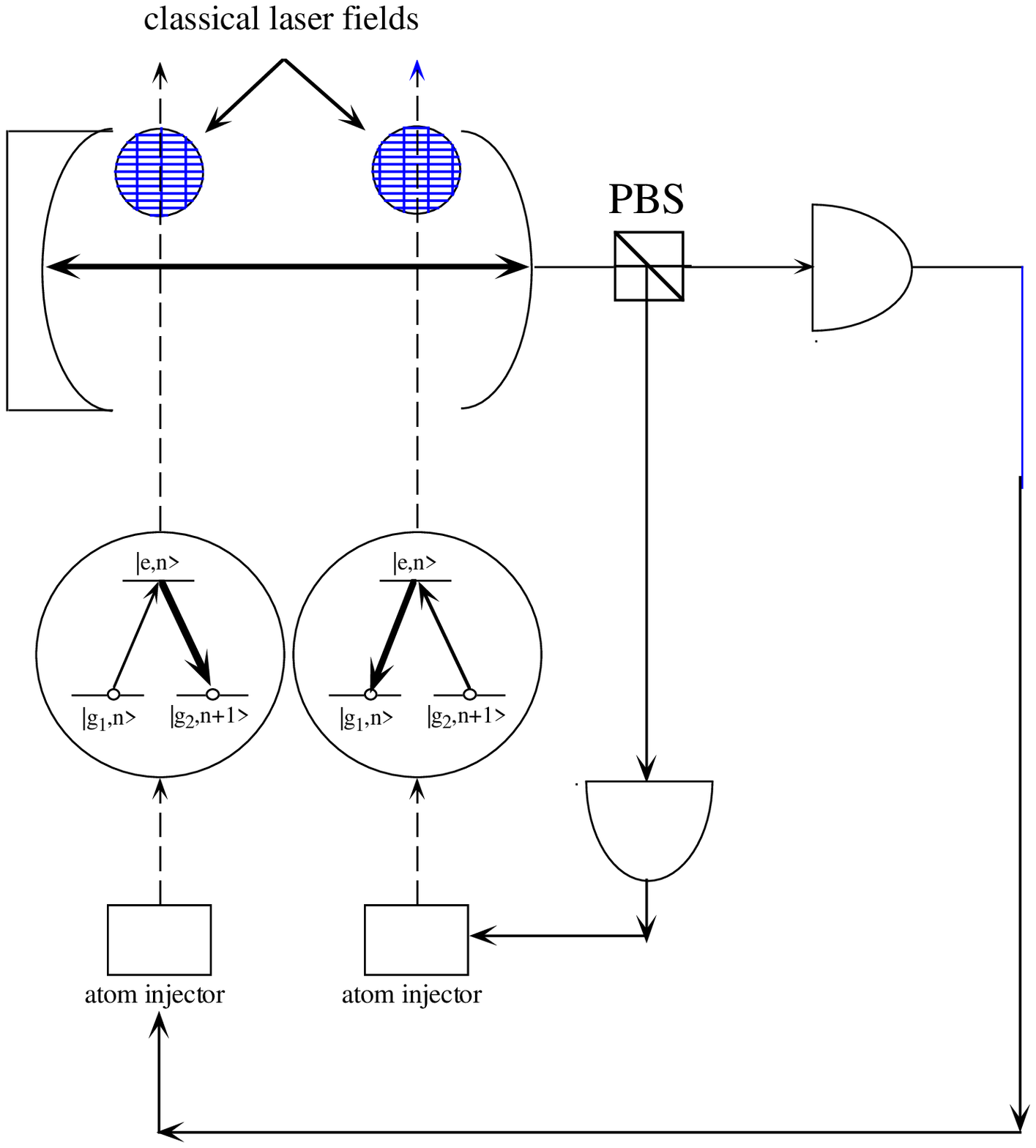,width=12cm}}
\caption{Adaptation of the feedback scheme of Fig.~1 to the 
polarization coding case; there is a feedback loop
for each circularly polarized mode and the two loops are separated by 
the polarized beam splitter PBS.}
\label{appa2}
\end{figure} 

For a quantitative characterization of how the feedback scheme is able 
to protect an initial pure state we study the fidelity $F(t)$
\begin{equation}
F(t) = {\rm Tr}\left\{\rho(0) \rho(t)\right\}
\label{fido}
\end{equation}
i.e., the overlap between the final and the initial state $\rho(0)$ 
after a time $t$. In general $0\leq F(t)\leq 1$. For an initially pure state
$|\psi
(0)\rangle$, $F(t)$ is in fact the probability to find the system in the 
initial state at a later
time. A decay to an asymptotic limit is given by the overlap $\langle \psi (0)
|\rho
(\infty)|\psi (0)\rangle$. 
Since the input state we seek to protect is unknown, the protection capabilities
of the feedback scheme are better characterized by the minimum fidelity,
i.e., the fidelity of Eq.~(\ref{fido}) minimized over all 
possible initial states. Moreover we shall consider a class of 
initial states more general than those of Eq.~(\ref{twoqub}), i.e., 
\begin{equation}
|\psi \rangle =  \alpha |n,m\rangle + \beta |m,n\rangle \;.
\label{twoqubgen}
\end{equation} 

Since the two polarized modes evolve independently, one has to solve 
the master equation (\ref{sqroot}) to calculate the fidelity. 
This can be done easily and one 
gets the following expression for the minimum fidelity
\begin{equation}
F_{min}(t)=\frac{1}{2}\left(
e^{-(1-\eta )\gamma t (n+m)}+e^{-\gamma t (n+m-2\eta \sqrt{n m})}\right) 
\;.
\label{minfid}
\end{equation}

In the absence of feedback ($\eta =0$), this expression becomes 
$F_{min}(t)=\exp\{-\gamma t(n+m)\}$ showing that in this case, the 
states  most robust against cavity damping are those with the smallest number 
of photons, $m+n=1$, i.e., the states of the form of Eq.~(\ref{twoqub}).
Moreover, in a typical quantum information processing situation, one has 
to consider small qubit ``storage'' times $t$ with respect to $\gamma 
^{-1}$ so to have reasonably small error probabilities in quantum 
information storage. Therefore the protection capability of an optical 
cavity with no feedback applied is described by 
\begin{equation}
F_{min}(t) = 1-\gamma t \;.
\label{nofed}
\end{equation}

If we now consider the situation in the presence of feedback 
(Eq.~(\ref{minfid})), it is possible to see that, for fixed, non-unit 
efficiency $\eta $, the best protected state are, as in the no-feedback 
case, the states with only one photon $\alpha |0,1\rangle+\beta 
|1,0\rangle$ and therefore the corresponding minimum fidelity for $\eta 
<1$ is given by
\begin{equation}
F_{min}(t)=\frac{1}{2}\left(
e^{-(1-\eta )\gamma t}+e^{-\gamma t }\right) \simeq 1-\gamma t 
\left(1-\frac{\eta }{2}\right) \;.
\label{minfid2}
\end{equation}
This shows that feedback increases the ``lifetime'' of a  
generic qubit state with respect to the no-feedback case, even if, in this 
non-ideal case, one has a scaling of the error probability by a 
factor $(1-\eta /2)$ only. 

It is interesting to consider the ideal case $\eta =1$, even if it is not 
realistic, because in this case the situation changes qualitatively. In 
fact Eq.~(\ref{minfid}) becomes
\begin{equation}
F_{min}(t)=\frac{1}{2}\left(
1+e^{-\gamma t (\sqrt{n}-\sqrt{m})^{2}}\right) 
\;.
\label{minfid3}
\end{equation}
so that it is easy to see that in this case it becomes convenient to work 
with a large number of photons and that the best protected qubit 
states are those of the form
\begin{equation}
|\psi \rangle =\alpha |n,n+1\rangle +\beta |n+1,n\rangle \;\;\;\;\; n \gg1
\label{larpho}
\end{equation}
whose corresponding minimum fidelity is
$$
F_{min}(t)\simeq  \frac{1}{2}\left(1+ 
e^{-\gamma t/4n}\right) \simeq 1-\frac{\gamma t}{8n} \;.
$$
Therefore, in the ideal photodetection case and using qubits of the form of 
(\ref{larpho}), the probability of errors in the storage of quantum 
information can be made arbitrarily small.

The feedback method proposed here to deal with decoherence in quantum 
information processing is different from most of the proposals made in 
this research field, which are based
on the so called quantum error correction codes \cite{error}. In our 
case, feedback allows a physical control of decoherence, through a 
continuous monitoring and eventual correction of the dynamics and in this 
sense our approach is similar in spirit to the approach of 
Ref.~\cite{pell,mabuchi}.
Quantum error correction is instead a way to 
use {\it software} to preserve linear superposition states. Essentially 
in these approaches the entangled superposition state of $l$ qubits is 
``encoded'' in a larger number 
of qubits $n$, so that, assuming that only a fraction of qubits $t/n$
decoheres, it is possible to reconstruct the original state with a 
suitable decoding procedure. However, due to existence of a lower (quantum 
Hamming) and an upper (quantum Gilbert-Varshamov) bound for the 
``code rate'' $l/n$ \cite{chia}, these quantum error correction codes can 
be applied and become efficient only at sufficiently small probability 
of error $t/n$. For this reason, even if under realistic conditions our
feedback scheme achieves only a moderate reduction of the error 
probability, it could be useful when used in {\it conjunction} with 
quantum error correction techniques. The feedback scheme would realize
the preliminary reduction of the error probability, which is necessary for 
an optimal implementation of efficient error correction schemes.

\section{Acknowledgments}
This work has been partially supported by the Istituto Nazionale Fisica 
della Materia (INFM) through the ``Progetto di Ricerca Avanzata INFM-CAT''.


\begin{thebibliography}{99}


\bibitem{bennet}C.H. Bennett, in {\it Quantum Communication, Computing 
and Measurement}, edited by O. Hirota, A.S. Holevo and C.M. Caves (Plenum 
Press, New York, 1997), pag. 25. 
\bibitem{eke}A. Ekert and R. Josza, Rev. Mod. Phys {\bf 68}, 733 (1996).
\bibitem{zur}W.H. Zurek, Phys. Today {\bf 44}(10), 36 (1991), and
references therein.
\bibitem{leg1}A.J. Leggett, in {\em Chance and Matter}
(Proceedings, 1986 Les Houches Summer School), ed. by J. Souletie,
J. Vannimenus, and R. Stora, (North Holland, Amsterdam, 1987),
pag. 395.
\bibitem{turchette}Q.A. Turchette, C.J. Hood, W. Lange, H. Mabuchi and 
H.J. Kimble, Phys. Rev. Lett. {\bf 75}, 4710 (1995).
\bibitem{wine1}C. Monroe, D.M. Meekhof, B.E. King, W.M. Itano and D.J. 
Wineland, Phys. Rev. Lett. {\bf 75}, 4714 (1995).  
\bibitem{prlno}D. Vitali, P. Tombesi, G.J. Milburn, Phys. Rev. Lett. 
{\bf 79}, 2442 (1997).
\bibitem{jmo}D. Vitali, P. Tombesi, G.J. Milburn, J. Mod. Opt. {\bf 
44}, 2033 (1997).
\bibitem{feed}H.M. Wiseman and G.J. Milburn, Phys. Rev. Lett.
{\bf 70}, 548 (1993); Phys. Rev. A {\bf 49}, 1350 (1994);
H.M. Wiseman, Phys. Rev. A {\bf 49}, 2133 (1994).
\bibitem{noi}P. Tombesi and D. Vitali, Phys. Rev. A {\bf 51},
4913 (1995); P. Goetsch, P. Tombesi and D. Vitali, Phys. Rev. A {\bf 54}, 
4519 (1996).
\bibitem{Carmichael}H.J. Carmichael, {\it An Open Systems Approach to Quantum 
Optics}, (Springer, Berlin, 1993). 
\bibitem{squee}P. Tombesi and D. Vitali, Phys. Rev. A {\bf 50}, 4253 
(1994).
\bibitem{qnoise}C.W. Gardiner, {\it Quantum Noise}, (Springer, 
Berlin, 1991). 
\bibitem{poyatos}J.F. Poyatos, J.I. Cirac and P. Zoller, Phys. Rev. Lett. 
{\bf 77}, 4728 (1996).
\bibitem{adia}A.S. Parkins, P. Marte, P. Zoller, O. Carnal and H.J. 
Kimble, Phys. Rev. A {\bf 51}, 1578 (1995) and references therein.
\bibitem{messiah}A. Messiah, {\it Quantum Mechanics} (North Holland, 
Amsterdam, 1962).
\bibitem{chim}J.R. Kuklinski, U. Gaubatz, F.T. Hioe and K. Bergmann, 
Phys. Rev. A {\bf 40}, 6741 (1990).
\bibitem{wise}H.M. Wiseman, Phys. Rev. A {\bf 51}, 2459 (1995).
\bibitem{fok}K. Vogel, V.M. Akulin and W.P. Schleich, Phys. Rev. Lett. 
{\bf 71}, 1816 (1993).
\bibitem{eberly}C.K. Law and J.H. Eberly, Phys. Rev. Lett. {\bf 
76}, 1055 (1996).
\bibitem{sten}S. Stenholm, Opt. Comm. {\bf 123}, 287 (1996); P. T\"orm\"a and 
S. Stenholm, Phys. Rev. A {\bf 54}, 4701 (1996).
\bibitem{error}E. Knill and R. Laflamme, Phys. Rev. A {\bf 55}, 900 (1997)
and references therein.
\bibitem{pell}T. Pellizzari, S.A. Gardiner, J.I. Cirac,
and P. Zoller, Phys. Rev. Lett. {\bf 75}, 3788 (1995).
\bibitem{mabuchi}H. Mabuchi and P. Zoller, 
Phys. Rev. Lett, {\bf 76}, 3108 (1996).
\bibitem{chia}A. Ekert and C. Macchiavello, Phys. Rev. Lett. 
{\bf 77}, 2585 (1996).

\end{thebibliography}
\end{document}